# Photoacoustic Imaging Based on AlN MF-PMUT with Broadened Bandwidth

Junxiang Cai†, Yiyun Wang†, Daohuai Jiang, Yuandong (Alex) Gu, Liang Lou,
Fei Gao* and Tao Wu*

*Abstract*—This paper reports an aluminum nitride (AlN) multi-frequency piezoelectric micromachined ultrasound transducers (MF-PMUT) array for photoacoustic (PA) imaging, where the broadened bandwidth is beneficial to improve imaging resolution. Specifically, PMUT based on micro-electromechanical systems (MEMS) technology is suitable for PA endoscopic imaging of blood vessels and bronchi due to its miniature size. More importantly, AlN is a non-toxic material, which makes it harmless for biomedical applications. In this work, a MF-PMUT array are designed and fabricated for PAI. The device's vibration mode impedance and bandwidth are analyzed. The MF-PMUT sensor provides a wider bandwidth (65%) signal detection, which increases the resolution of PAI compared with traditional PMUT. We conduct an experiment on agar sample to present sensor's performance in images' axial resolution.

## I. INTRODUCTION

Micromachined ultrasonic transducers based on piezoelectric (PMUT) have attracted attentions to the biomedical sensing and imaging, due to their low-power consumption and small size [1]. Biomedical imaging system, such as photoacoustic imaging (PAI) system, tend to be compact and flexible in recent decades [2]–[4]. Due to its nice thermal and chemical stability, aluminum nitride (AlN) thin film based MF-PMUT has raised attentions and is applied for miniaturizing the photoacoustic imaging system [5]–[7].

PMUT at fundamental resonance has the property of high sensitivity and narrow frequency bandwidth [5], [8], [9]. It would result in the image's insufficient axial resolution. The ultrasonic transducer's frequency bandwidth is proportional to the PA signal's full width of half maximum (FWHM), which is the main indicator for evaluating the image's axial resolution [10], [11]. It is desired that a PAI integration system based on AlN multi-frequency PMUT with a relatively broad frequency bandwidth. Such universal system can be applied to various biomedical imaging scenarios. Thus, increasing PMUT's bandwidth is a concerned topic. According to the working modes, PMUTs are usually ranged into two large categories: flexural vibration mode (FVM) and thickness extension mode (TEM). Between two modes, FVM-PMUTs' device fabrication is more compatible with the modern fabrication process of the complementary metal-oxide semiconductor (CMOS). The advantage in fabrication makes it more possible to produce monolithic transducer chips [7], [12], [13]. They also have the lower acoustic impedance and are easier to integrate multiple frequency bands.

In PAI, both lateral and axial resolutions are concerned, affecting the quality of images[14]–[16]. The lateral resolution is determined by the overlap of optical excitation and ultrasonic detection [16]. While, the axial resolution originates from the PA signal's full width of half maximum (FWHM). It is proportional to the acoustic detector's bandwidth. Typically, PA signal has a short pulse profile with a wide bandwidth [15], [17]–[19]. It is highly required to use a wide bandwidth acoustic sensor to acquire PA signal with high fidelity. Therefore, increasing the bandwidth of the transducer is highly essential to improve the resolution of PAI.

In this manuscript, we propose the photoacoustic system based on a bandwidth-expanding AlN FVM-PMUT. With the PMUT's broadened bandwidth, the system is available for imaging tissues with better axial resolution.

## II. PIEZOELECTRIC MICROMACHINED ULTRASOUND TRANSDUCER

### A. MF-PMUT design and fabrication

Fig. 1 shows the MF-PMUT device. The MF-PMUT device is fabricated on SOI (Silicon-On-Insulator) wafer shows in Fig. 1(b), and AlN thin film is its piezoelectric layer. Molybdenum (Mo) layers are used as the top and bottom electrodes to collect the charge of AlN piezoelectric layer. The Mo electrodes can also drive the AlN piezoelectric layer by applying voltage. Through two-step etching, material aluminum (Al) can connect the bottom electrode and the top electrode to the surface thru wire bounding.

The manuscript is received at xxx. This work was supported in part by the National Natural Science Foundation of China (61874073), Natural Science Foundation of Shanghai (19ZR1477000) and Pujiang Talent Program (19PJ1432300). *(Corresponding author: Fei Gao and Tao Wu).* The first two authors contribute equally.

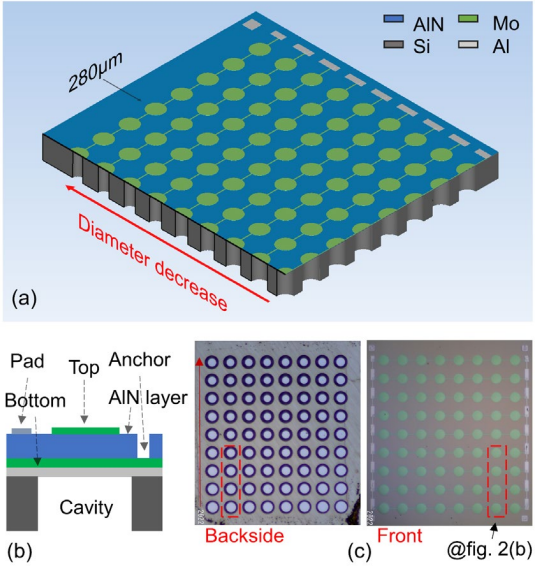

**Fig. 1.** (a) The 3D model of MF-PMUT; (b) Sectional view of PMUT's cell; (c) Optical image of the MF-PMUT.

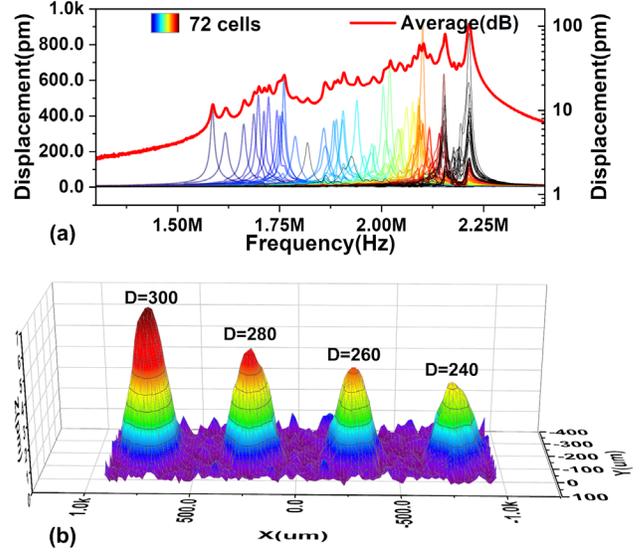

**Fig. 2.** (a) Frequency response simulation of MF-PMUT; (b) 72 cells' LDV measurement of MF-PMUT in air; (c) Mode shape of selected 4 cells of MF-PMUT.

Finally, a backside DRIE process creates cavities, which defines the effective flexural membrane diameter of the MF-PMUT cell. The 3D model of the MF-PMUT array is shown in Fig. 1(a). The MF-PMUT array contains 8 rows and 9 columns, a total of 72 cells, and all the cells are connected with top electrode. The diameter of MF-PMUT gradually decreases along the row direction. The maximum is 300μm and the minimum is 180μm. Fig. 1(c) shows the optical front and back image of fabricated MF-PMUT.

*B. MF-PMUT device characterization*

The MF-PMUT is analyzed by laser doppler vibrometer (LDV, MSA-600 Micro System Analyzer, Polytec, Germany), impedance Analyzer and pulse-echo experiment. Fig. 2(b) shows the LDV measurement result of all the 72 cells. The PMUT array was electrically driven at 1 $V_{pp}$ with a 1 MHz – 3 MHz sweeping frequency in air. The fundamental resonance of 72 cells of MF-PMUT is in the range of 1.6 MHz(minimum)-2.2 MHz(maximum). Due to the deviation of the fabrication process and layout design, the resonant frequency of MF-PMUT is close to uniform distribution in this range, instead of having only 9 groups. Fig 2(b) shows the 4 cells' mode shape measured by LDV, and the four scanned cells correspond to the cell in the dotted rectangle. The cells was driven at non-resonant frequency, the larger diameter cells have higher amplitudes.

Fig. 3(a) shows the impedance measurement result of MF-PMUT array. The device was tested in air. The black curve is impedance absolute value and the red curve is phase. As shown in the Fig. 3(a), the phase distortion appears in the frequency range of 1.6 MHz-2.2 MHz, which corresponds to LDV measurement. Fig. 3(c) shows the pulse-echo experiment result of MF-PMUT array. The experiments were carried out in oil, the device was connected with manually controlled pulsed-receivers (Panametrics 5072PR, Panametrics, American). As it shown in the result, the peak frequency $f_p$ of MF-PMUT in mineral oil is about 1.89 MHz, the low frequency $f_l$ (6dB lower than peak in echo measurement, smaller than $f_p$) is 1.38 MHz and the high frequency $f_h$ (6dB lower than peak in echo measurement, bigger than $f_p$) is 2.71 MHz The bandwidth of MF-PMUT is about 65%.

Through device characterization, we can find that individual resonances of different diameter cells can fuse together when the cells are grouped together and form an extended bandwidth. The bandwidth of this MF-PMUT is significantly higher than single AlN PMUT array [20], which benefit improve imaging resolution of PAI. What's more, the bandwidth of MF-PMUT can be higher with further design.

### III. EXPERIMENT

*A. System Setup*

Here, we present the experimental setup of the photoacoustic sensing and imaging based on the AlN MF-PMUT which has been broadened its frequency bandwidth. The designed MF-PMUT is integrated on the PCB board. Beams of 532 nm pulsed lasers (DPS-532-A, CNI Laser) travel through the tiny hole on the PCB board and excite the phantom in the oil tank to generate photoacoustic signals. The PA signals are then received by the PMUT. They pass through a high-pass filter to reduce noise below 20 kHz, and are amplified by a charge amplifier. The signals are finally acquired by the oscilloscope (DPO5204B, Tektronix Inc.) and transmitted to the computer. The sample is scanned point by point in the X and Y axes, and the scanning step is 0.5 mm.

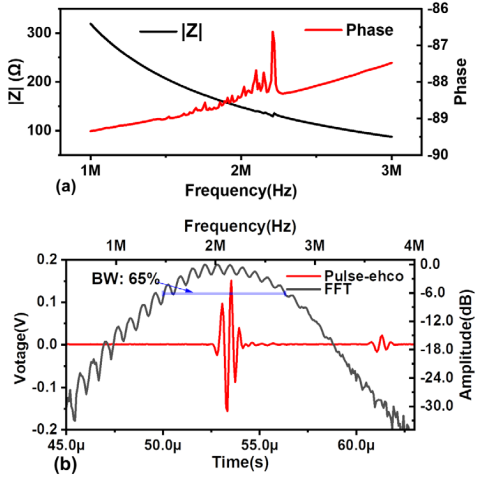

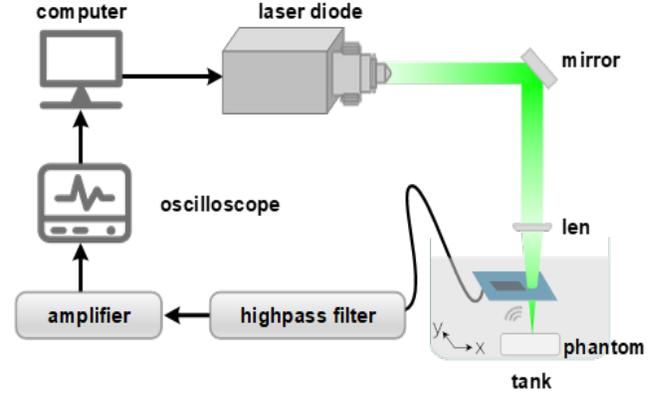

**Fig. 3.** (a) Impedance of MF-PMUT in air; (b)-6dB bandwidth of MF-PMUT measured by pulse-echo in oil.

**Fig. 4.** The system setup for the photoacoustic imaging experiment. The tank is filled with mineral oil.

*B. Results and Analysis*

We conduct the experiment on a block of agar phantom that is inserted with three pseudo-parallel pencil leads at different depths (Fig. 5(a)). Pencil leads' diameter is 0.5 mm. Here, we first analyze PMUT's performance in the aspect of PA signal waveform. As described above, the designed MF-PMUT is broadened bandwidth based on the original single-frequency PMUT. According to our previous work [20], single-frequency PMUT's PA signal has large oscillation. This kind of MF-PMUT is not possible to conduct PA imaging experiments that require depth information in the practical scenarios. Compared with the single-frequency PMUT's performance at fundamental resonance, the proposed MF-PMUT performs the neglectable oscillation in PA signals (Fig. 5(b)), which ensure relatively high axial resolution and the transducer's ability of visualizing depth information. Compared with the high-order resonance mode in our another work [21], the performances in these two kinds of modes and devices have similar axial resolution. But the device presented here has larger PA signal amplitude under the same amplification setup. It shows that the MF-PMUT has much better sensitivity. Fig. 5(b) also shows that the latency between the PA signals of the adjacent two pencil leads is about 2 us. It indicates 3 mm depth difference, which is close to the actual measurements.

Then, we visualize the photoacoustic results in the cross-sectional images. Fig. 5(c) displays the averaged results in the xz-plane cross section. Since the pencil leads are not strictly paralleled, we average 6 imaging results that include the lead's information for each pencil lead. We may observe the graphs and find relatively thicker pencil leads than that in Fig. 5(a), which is due to the average method. The results in x-z plane show that three pencil leads are placed at different depths. It can be observed clearer from the y-z plane cross section at x = 9.5 mm in Fig. 5(d).

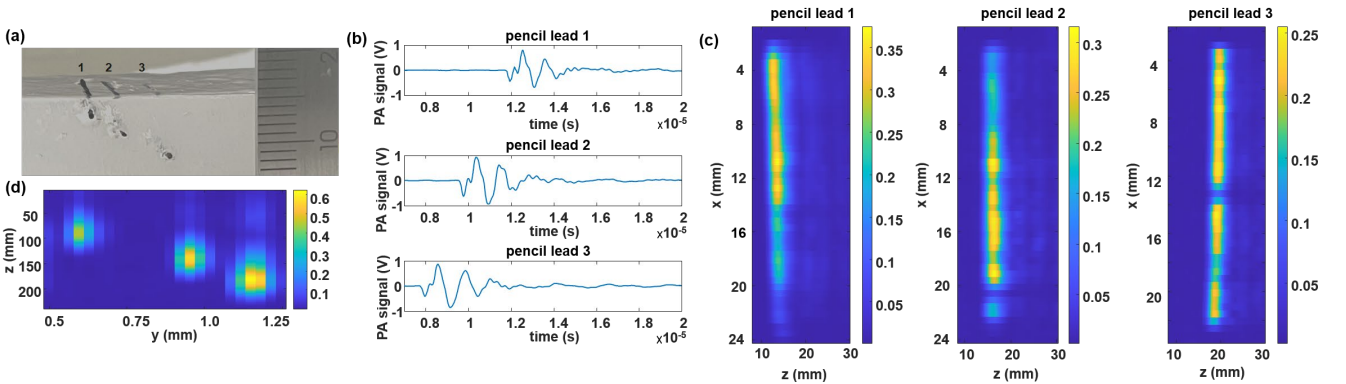

**Fig. 5.** (a). The agar phantom inserted with three pencil leads at different depths. The diameter of the pencil leads is 0.5 mm. (b). Three pencil leads' photoacoustic signals at x = 11 mm. (c). Leads' cross-sectional PA images on the xz-plane. (d). The leads' PA images of yz-plane cross section at x = 9.5 mm.

## IV. Conclusion

Based on the critical need of increasing PMUT's bandwidth, we propose an AlN multi-frequency PMUT sensor for photoacoustic imaging system. With the broadened bandwidth, we are able to provide depth information in the imaging process, compared with the single-frequency PMUT at fundamental resonance mode. Also, our experiment results also show that the sensor's sensitivity is better than that of single-frequency PMUT at high-order resonance mode. Thus, the designed PMUT might be suitable for more practical and universal application scenarios.

## Acknowledgment

The authors appreciate the MF-PMUT fabrication support from SITRI and ShanghaiTech Quantum Device Lab (SQDL).